\documentclass[floatfix,aps,twocolumn,superscriptaddress,showpacs,preprintnumbers,amsmath,amssymb]{revtex4}

\usepackage[applemac]{inputenc}
\usepackage[T1]{fontenc}
\usepackage{amsfonts, amssymb, amsmath, dsfont}
\usepackage{graphicx}
\usepackage{float}
\usepackage{ulem}
\usepackage{color}

\newcommand{\bPhi}{\mathbf{\Phi}}

\newcommand{\bE}{\mathbf{E}}

\newcommand{\calG}{\mathcal{G}}

\newcommand{\bG}{\mathbf{G}}

\newcommand{\br}{\mathbf{r}}

\newcommand{\bp}{\mathbf{p}}

\newcommand{\bu}{\mathbf{u}}

\newcommand{\be}{\mathbf{e}}

\def\d{\mathrm{d}}
\def\eps{\varepsilon}
\def\Re{\mathrm{Re}}
\def\Im{\mathrm{Im}}

\newcommand{\ImG}{\mathrm{Im} G}

\def\ge{|\mathrm{g}\mathrm{e}\rangle}
\def\eg{|\mathrm{e}\mathrm{g}\rangle}
\def\ee{|\mathrm{e}\mathrm{e}\rangle}

\renewcommand{\emph}{\textit}

\begin{document}

\title{Quantum coherence of light emitted by two single-photon sources \\ in a structured environment}

\author{Antoine Canaguier-Durand}
\affiliation{ESPCI ParisTech, PSL Research University, CNRS, Institut Langevin, 1 rue Jussieu, F-75005, Paris, France.}
\affiliation{Laboratoire Kastler Brossel, UPMC Sorbonne Universit\'es, CNRS, ENS, PSL Research University, Campus Jussieu, F-75252 Paris, France.}
\author{Rémi Carminati}
\altaffiliation{remi.carminati@espci.fr}
\affiliation{ESPCI ParisTech, PSL Research University, CNRS, Institut Langevin, 1 rue Jussieu, F-75005, Paris, France.}


\begin{abstract} 
We develop a theoretical framework for the analysis of the quantum coherence of light emitted by two independent single-photon sources in an arbitrary environment. The theory provides design rules for the control of the degree of quantum coherence, in terms of classical quantities widely used in nanophotonics. As an important example, we derive generalized conditions to generate superradiant and subradiant states of the emitters, and demonstrate the ability of a structured environment to induce long-range quantum coherence. These results should have broad applications in quantum nanophotonics, and for the sensing of fluorescent sources in complex environments. 
\end{abstract}

\maketitle

\emph{Introduction - }
The ability to change the dynamics of quantum emitters by structuring the electromagnetic environment has been the early motivation of cavity quantum electrodynamics \cite{purcell1946spontaneous,drexhage1970influence,haroche2013nobel}, has inspired the development of photonic crystals \cite{yablonovitch1987inhibited} and has become a major goal in nanophotonics, in which cavity or antennas concepts have been downscaled to the nanometer range~\cite{novotny2012principles}. Beyond changing the dynamics of isolated emitters, which is chiefly driven by the local density of states (LDOS), controlling the interactions among an ensemble of quantum emitters with nanostructures is a central issue in the emerging field of quantum nanophotonics \cite{lodahl2015interfacing,dzsotjan2010quantum,martin2010resonance,gonzalez2011entanglement,tame2013quantum,goban2015superradiance}.
Using a nanostructured environment to drive the quantum coherence of the light emitted by two (or more) single-photon sources would be a major step forwards in many areas, including the treatment of quantum information in integrated photonics \cite{kimble2008quantum}, or the control of collective emission~\cite{pustovit2009cooperative,oppel2014directional,wiegner2015simulating} and absorption for the design of novel efficient light sources and absorbers. Establishing a clear connection between the degree of quantum coherence of the emitted light and the local environment of the emitters could also stimulate new strategies for the detection of sources in complex media (such as biological tissues), along the lines initiated in Ref.~\cite{carminati2015speckle} for classical sources. It would also help the understanding of the role of quantum coherence in photosynthetic light-harvesting systems, an issue of high current interest~\cite{engel2007evidence,collini2010coherently,anna2014little}. 

In this Letter, we study the second-order quantum coherence of light emitted by two independent single-photon sources in an arbitrary electromagnetic environment. We establish a general theoretical framework, in which design rules for the control of the degree of quantum coherence naturally emerge. As an important example, we derive the conditions for the observation of subradiant and superradiant states. In the case of a detection integrated over all output channels, the photodetection correlation functions are expressed in terms of the local and cross densities of states of the electromagnetic field, allowing a direct connection to classical quantities widely used in nanophotonics. 

\emph{Geometry and photodetection signals - }
We consider two single-photon emitters located at positions $\br_1$ and $\br_2$ in an arbitrary environment.
The emitters are modeled as quantum two-level systems, with the same transition energy $\hbar \omega$ and transition dipoles $\bp_1 = p_1 \left( \sigma_1^+ + \sigma_1^- \right) \bu_1$ and $\bp_2=p_2  \left( \sigma_2^+ + \sigma_2^- \right) \bu_2$, where $\sigma^\pm$ are atomic raising and lowering operators, and $\bu_1$ and $\bu_2$ are fixed unit vectors. The two emitters are assumed to be noninteracting, so that the quantum states of the ensemble are product states (such as $\eg =| \mathrm{e} \rangle_1 \otimes | \mathrm{g} \rangle_2$ when emitter 1 is excited and emitter 2 is in the ground state).

\begin{figure}[htbp]
\centering{
\vspace{0.cm}
\includegraphics[width=0.35\textwidth]{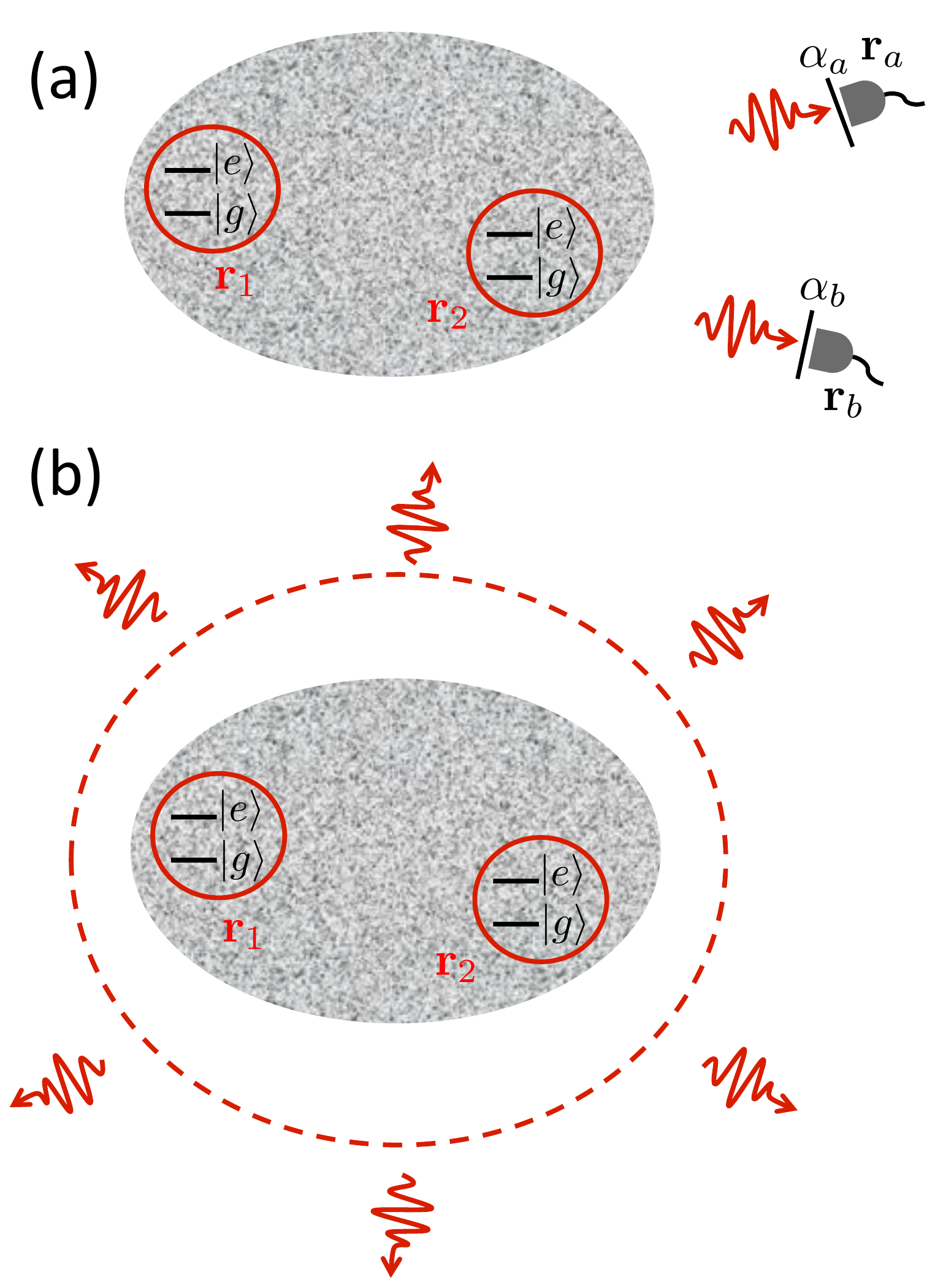}
\vspace{0.cm}
}
\caption{(Color online) Photodetection geometry. (a): Two quantum emitters are located at positions $\br_1$ and $\br_2$ inside
a structured medium, and polarized detectors are located at positions $\br_a$ and $\br_b$. (b): Integrated photodection where the photodetection signals are integrated over all directions and polarizations.}
 \label{figure1}
\end{figure}

We first assume a photodetection scheme using two detectors at positions $\br_a$ and $\br_b$, as sketched in Fig.~\ref{figure1}(a). The detectors select photons with polarization states $\alpha_a$ and $\alpha_b$, corresponding to the projections of the electric field along the unit vectors $\be_a$ and $\be_b$. 
The positive-frequency component of the electric field operator can be connected to the source operators using the electric Green function:
\begin{align}
\bE^{(+)} (\br) = \mu_0 \omega^2  \left [  p_1 \sigma_1^- \bG (\br, \br_1) \cdot \bu_1 \right. \nonumber \\  \left. + p_2 \sigma_2^- \bG(\br,\br_2) \cdot\bu_2\right ] .
\end{align} 
Here the Green function $\bG$ is evaluated at the angular frequency $\omega$, and contains all the information about the surrounding environment~\cite{novotny2012principles}. For simplicity we introduce the simplified notations $E_a=\be_a \cdot \bE^{(+)}(\br_a)$ and $G_{ai}=\be_a \cdot \bG(\br_a,\br_i) \cdot \bu_i$ that describe quantities measured at detector $a$ (and similarly for detector $b$). 
The photodetection of one photon at position $\br_a$, with polarization state $\alpha_a$, is described by the operator $\bPhi_1(\br_a, \alpha_a) = E_a^\dagger E_a$. In the particular case of two emitters in the excited state, the probability to detect one photon at position $\br_a$, with polarization $\alpha_a$, is given by the expectation value 
\begin{align}\label{phi1_ee}
 \langle  \bPhi_1 (\br_a, \alpha_a) \rangle &=  \left( \mu_0 \omega^2 \right)^2 \left[  |p_1G_{a1} |^2+ |p_2 G_{a2} |^2 \right]  
\end{align}
that includes two independent contributions from each emitter, as expected for independent sources.
The photodetection of two photoncs at positions $(\br_a, \br_b)$, with respective polarizations $(\alpha_a,\alpha_b)$, is described by the operator $\bPhi_2 (\br_a,\alpha_a,\br_b,\alpha_b) =E_a^\dagger E_b^\dagger E_b E_a$. 
When the two emitters are in the excited state, its expectation value
\begin{align}\label{phi2_ee}
\langle \bPhi_2 (\br_a,\alpha_a,\br_b,\alpha_b) \rangle  = 
 \left( \mu_0 \omega^2 \right)^4 |p_1 p_2|^2 \nonumber \\ \times \left| G_{a1} G_{b2} + G_{a2} G_{b1} \right|^2
\end{align}
gives the joint probability to detect one photon at position $\br_a$ with polarization $\alpha_a$ and one photon at position $\br_b$ with polarization $\alpha_b$.
This expression differs from a simple product of single photodetection probabilities, and is thus affected by interferences between the contributions of the two sources.

As a measure of the degree of quantum coherence of the emitted light, we introduce the second-order correlation factor \cite{glauber1963quantum}
\begin{align}\label{g2_quant}
&g^{(2)}(\br_a,\alpha_a,\br_b,\alpha_b) =  \frac{\langle \bPhi_2(\br_a,\alpha_a,\br_b,\alpha_b) \rangle}{\langle \bPhi_1(\br_a,\alpha_a )\rangle \cdot \langle \bPhi_1(\br_b,\alpha_b )\rangle} \nonumber \\
& ~ ~ = \frac{ |p_1 p_2|^2 |G_{a1} G_{b2} + G_{a2} G_{b1}|^2}{\left( |p_1 G_{a1}|^2 + |p_2 G_{a2} |^2 \right) \left( |p_1 G_{b1}|^2 + |p_2 G_{b2} |^2 \right)} ~ .
\end{align}
One can easily verify that $0\leq g^{(2)}(\br_a,\alpha_a,\br_b,\alpha_b) \leq 1$, meaning that for any positions of the detectors, the single-photon nature of the quantum emitters yields an antibunching behavior \cite{kimble1977photon,messin2001bunching}, regardless of the photonic environment [{\it i.e.} whatever the value of the Green functions in Eq.~(\ref{g2_quant})]. This result is different from that derived for incoherent classical emitters in Ref.~\cite{carminati2015speckle}, where an autocorrelation factor (for $\br_a=\br_b$ and $\alpha_a=\alpha_b$) with values between 1 and $3/2$ was obtained. This shows that the second-order correlation factor can reveal the quantum nature of the emitters, independently on the photonic environment. 

From Eq.~(\ref{g2_quant}), it is interesting to look for the conditions under which $g^{(2)}$ reaches its extremum values. One gets a maximum correlation $g^{(2)}=1$ when $ \left| p_1 \right|^2 G_{a1} G_{b1}^* =  \left| p_2 \right|^2  G_{a2} G_{b2}^*$, which can be split into two conditions on amplitudes and phases:
\begin{align}
 \left| p_1 \right|^2 \left| G_{a1} G_{b1} \right| &=  \left| p_2 \right|^2 \left| G_{a2} G_{b2} \right| \label{ampl_cond} \\
 \mathrm{arg} \left( G_{a1} G_{b2} \right) &=  \mathrm{arg} \left( G_{a2} G_{b1} \right) ~ . \label{phase_cond}
\end{align} 
The first condition (\ref{ampl_cond}) states that the efficiency to reach the two photodetectors has to be the same for both emitters (each side can be understood as a geometrical average of the probabilities to send a photon to each detector). Interestingly, this condition can be fulfilled even in an asymmetric configuration, for instance with the first emitter well connected to $\br_a$ and the second emitter well connected to $\br_b$. The second condition (\ref{phase_cond}) states that for the two possible scenarios giving rise to two measured photons, the accumulated phase shift must be the same in order to generate constructive interferences between the contributions of the two sources. 
Conversely, the condition to get $g^{(2)}=0$ is
\begin{align}
G_{a1} G_{b2} + G_{a2} G_{b1}  = 0 \label{cond_sub} 
\end{align} 
and does not depend on the amplitudes $|p_1|$ and $|p_2|$ of the two emitters. Indeed the condition simply states that the two possible scenarios for double photodetection must have the same amplitude and opposite phase to reach destructive interferences. 

\emph{Detection-induced coherence - }
The conditions maximizing or minimizing the degree of quantum coherence can be understood from a different perspective, by studying more specifically the correlations between the two photodetection processes. This provides an interpretation of the appearance of quantum coherence, and useful rules for the engineering of the photonic environment in order to generate superradiant and subradiant states of the two quantum emitters.

Since $\langle \bPhi_2 (\br_a,\alpha_a,\br_b,\alpha_b)\rangle$ is the joint probability to detect one photon at $\br_a$ with polarization $\alpha_a$ and one photon at $\br_b$ with polarization $\alpha_b$, starting from the state $\ee$ with two excitations, it can be rewritten as the product of the probability to detect one photon at $\br_a$ with polarization $\alpha_a$ by the conditional probability to detect one photon at $\br_b$ with polarization $\alpha_b$, knowing that the first photon has already been detected. In terms of expectation values, this reads
\begin{align}\label{decompo_phi2}
\langle \bPhi_2 (\br_a,\alpha_a,\br_b,\alpha_b)\rangle = \langle \bPhi_1 (\br_a,\alpha_a)\rangle \cdot  \langle \Psi_a | \bPhi_1 (\br_b,\alpha_b) | \Psi_a \rangle
\end{align}
 where $| \Psi_a \rangle$ is the state of the emitters after the measurement of the first photon. The projection realized by this first photodetection on the state $\ee$ yields $| \Psi_a \rangle \propto E_a \ee$, which after normalization leads to
 \begin{align}\label{projected_state}
| \Psi_a \rangle = \frac{p_1 G_{a1} \ge + p_2 G_{a2} \eg}{\sqrt{|p_1 G_{a1}|^2 + |p_2 G_{a2}|^2}} .
 \end{align}
This state is a superposition of two states with only one emitter in the excited state, with amplitudes and phases determined by the transition dipole amplitudes $p_1$ and $p_2$, and by the propagation from each emitter to the position of the photodetector at $\br_a$ (described by the Green functions $G_{a1}$ and $G_{a2}$). Starting from the state $\ee$, the first photodetection event has generated correlations between the two emitters, since it is not possible to know which source emitted the measured photon. 
Note that the superposition can be substantially unbalanced, due to different transition dipoles or propagator (Green's function) weights. The correlation factor defined in Eq.~(\ref{g2_quant}) can then be rewritten as 
\begin{align}\label{g2_interpretation}
g^{(2)}(\br_a,\alpha_a,\br_b,\alpha_b) = \frac{  \langle \Psi_a | \bPhi_1 (\br_b,\alpha_b) | \Psi_a \rangle}{ \langle \mathrm{ee} | \bPhi_1(\br_b,\alpha_b) \ee} \leq 1 
\end{align}
which enables to give a physical picture for the antibunching behavior. Indeed, the ratio in Eq.~(\ref{g2_interpretation}) now reads as a measure of the constraints induce by the first measurement on the second detection. There are two distinct constraints: First, the loss of one excitation reduces the expectation value for the second photodetection event; Second, the first measurement induces coherence between the two sources, which produces constructive or destructive interferences influencing the second photodetection. Getting a ratio lower than one means that constructive interferences cannot overtake the loss of one excitation. The first measurement thus always reduces (or keeps unchanged, at best) the probability to detect a photon at $\br_b$ with polarization $\alpha_b$, whatever the relative amplitudes and phases in the superposition state~(\ref{projected_state}) produced by the first photodetection. 

In this framework, the conditions (\ref{ampl_cond}) and (\ref{phase_cond}) leading to the maximum correlation factor $g^{(2)}=1$ can be understood as requirements for the coherence between the emitters to generate perfect constructive interferences. When these conditions are fulfilled, the loss of one excitation is completely compensated by the optimal correlation between the sources. This is the mechanism at the origin of the phenomenon of superradiance~\cite{gross1982superradiance,scheibner2007superradiance}, that in free space is observed only for emitters in close proximity.
Equations~(\ref{ampl_cond}) and (\ref{phase_cond}) actually provide generalized conditions to generate a superradiant state for emitters in an arbitrary photonic environment. In particular, superradiance can be obtained for distant emitters, provided that the structure of the photonic modes (described by the Green function in our formalism) permits to satisfy these two conditions. Similarly, the condition (\ref{cond_sub}) leading to $g^{(2)}=0$ ensures that destructive interferences are maximized. Consequently, the measurement-induced coherence between the sources suppresses emission towards the second photodetector. The system can be considered in a subradiant state, as the emission vanishes after the first photodetection event. 
In summary, conditions (\ref{ampl_cond}), (\ref{phase_cond}) and (\ref{cond_sub}) provide rules to engineer the photonic environment (the Green function) in order to control the degree of quantum coherence of single-photon emitters, up to the generation of superradiant and subradiant states.

\emph{Photodetection over all output channels - }
We now define the operators obtained after integration of the single and double photodetection operators over all possible directions and polarizations (or more generally over all possible output channels, {\it e.g.}, in a waveguide or cavity geometry):
\begin{align}
P_1&=\frac{\eps_0 c}{2} \int_{S_a} \d \br_a  \sum_{\alpha_a} \bPhi_1 (\br_a,\alpha_a)\\
P_2&=  \left( \frac{\eps_0 c}{2} \right)^2 \int_{S_a} \d \br_a  \int_{S_b} \d \br_b \sum_{\alpha_a,\alpha_b} \bPhi_2 (\br_a,\alpha_a,\br_b,\alpha_b) ~ .
\end{align}
The prefactors are used to define observables corresponding to radiated power. These operators involve angular integrations of products of two Green functions, that simplifies into imaginary parts of Green functions in the case of a {\it non-absorbing medium} (a similar calculation with classical sources is found in Ref.~\cite{carminati2015speckle}). Assuming the two emitters in the excited state, the probabilities to detect one or two photons over all output channels take the simple form
\begin{align}
\langle P_1 \rangle &=  \frac{\mu_0 \omega^3}{2} \left( |p_1|^2 \ImG_{11} + |p_2|^2 \ImG_{22} \right) \label{P1_ee}  \\
\langle P_2 \rangle &=  \frac{\mu_0^2 \omega^6}{2}  |p_1 p_2|^2 \left [\ImG_{11}\ImG_{22} + (\ImG_{12})^2 \right] \label{P2_ee}
\end{align}
where we have used the simplified notation $\ImG_{jk}= \bu_j \cdot \Im[\bG(\br_j,\br_k)] \cdot \bu_k$, and the equality $\ImG_{12}=\ImG_{21}$ that is a consequence
of reciprocity. The one-point imaginary part of the Green function $\ImG_{jj}$ is proportional to the local density of states (LDOS), that counts the contribution of
modes at a given point $\br_j$, while the two-point imaginary part $\ImG_{jk}$ is proportional to the cross density of states (CDOS), that describes intrinsic spatial coherence between the points $\br_j$ and $\br_k$~\cite{caze2013spatial}.
The one-photon detection probability (\ref{P1_ee}) can be split into two independent components relative to each emitter, with weights proportional to the corresponding LDOS, as expected. Conversely, the two-photon detection probability (\ref{P2_ee}) contains two different contributions:  While the term with a product of LDOSs describes the emission of one photon by each dipole without interaction, the product of CDOSs accounts for interferences between the two emission processes. 

A generalized correlation factor $\calG^{(2)}= \langle P_2 \rangle/\langle P_1 \rangle^2$ can be defined for measurements integrated overall output channels, and reads as
\begin{align}\label{G2_quant}
\calG^{(2)} =  \frac{2 |p_1 p_2|^2 \left [ \ImG_{11} \ImG_{22}+ (\ImG_{12})^2 \right ]}{\left( |p_1|^2 \ImG_{11} + |p_2|^2 \ImG_{22} \right)^2} ~ .
\end{align}  
Since the inequality $\left| \ImG_{12} \right| \leq \sqrt{ \ImG_{11}} \sqrt{\ImG_{22}}$ is satisfied (see the Appendix for a derivation), we have $0\leq \calG^{(2)} \leq 1$, showing that the antibunching behavior is conserved after integration over all output channels. This is also different from the classical case, for which the correlation factor (defined as the ratio between averaged emitted power and power fluctuations) takes values between 1 and $3/2$~\cite{carminati2015speckle}.

Similarly to the case of local photodetections, conditions for the generation of superradiant and subradiant states can be derived for the output-channels-integrated photodetection scheme by looking at extrema of the correlation factor $\calG^{(2)}$. Superradiant states, that enable $\calG^{(2)}=1$, are obtained under the conditions 
\begin{align}
  |p_1|^2 \ImG_{11} &= |p_2|^2 \ImG_{22} \label{G2cond_opt_1} \\
   (\ImG_{12})^2 &= \ImG_{11} \ImG_{22} \label{G2cond_opt_2}
\end{align} 
showing that the two emitters must have the same emissive power, and the CDOS connecting their positions has to be maximum (meaning that the two sources have to be
highly connected by the photonic modes supported by the structured environment~\cite{caze2013spatial}). Although in free space this second condition is only satisfied for sources separated by a subwavelength distances, in a structured environment this range can in principle be arbitrary large.
Conversely, subradiant states producing $\calG^{(2)}=0$ are generated when $|p_1|^2 \ImG_{11}$ and $ |p_2|^2 \ImG_{22} $ have very different magnitudes, and $\ImG_{12} \simeq 0$. These conditions mean that the two sources must have very different emissive powers, and must be weakly connected by the mode structure of the photonic environment. Unlike in free space, these conditions can be satisfied even for emitters at subwavelength distance.

\begin{figure}[htbp]
\centering{
\vspace{0.cm}
\includegraphics[width=0.4\textwidth]{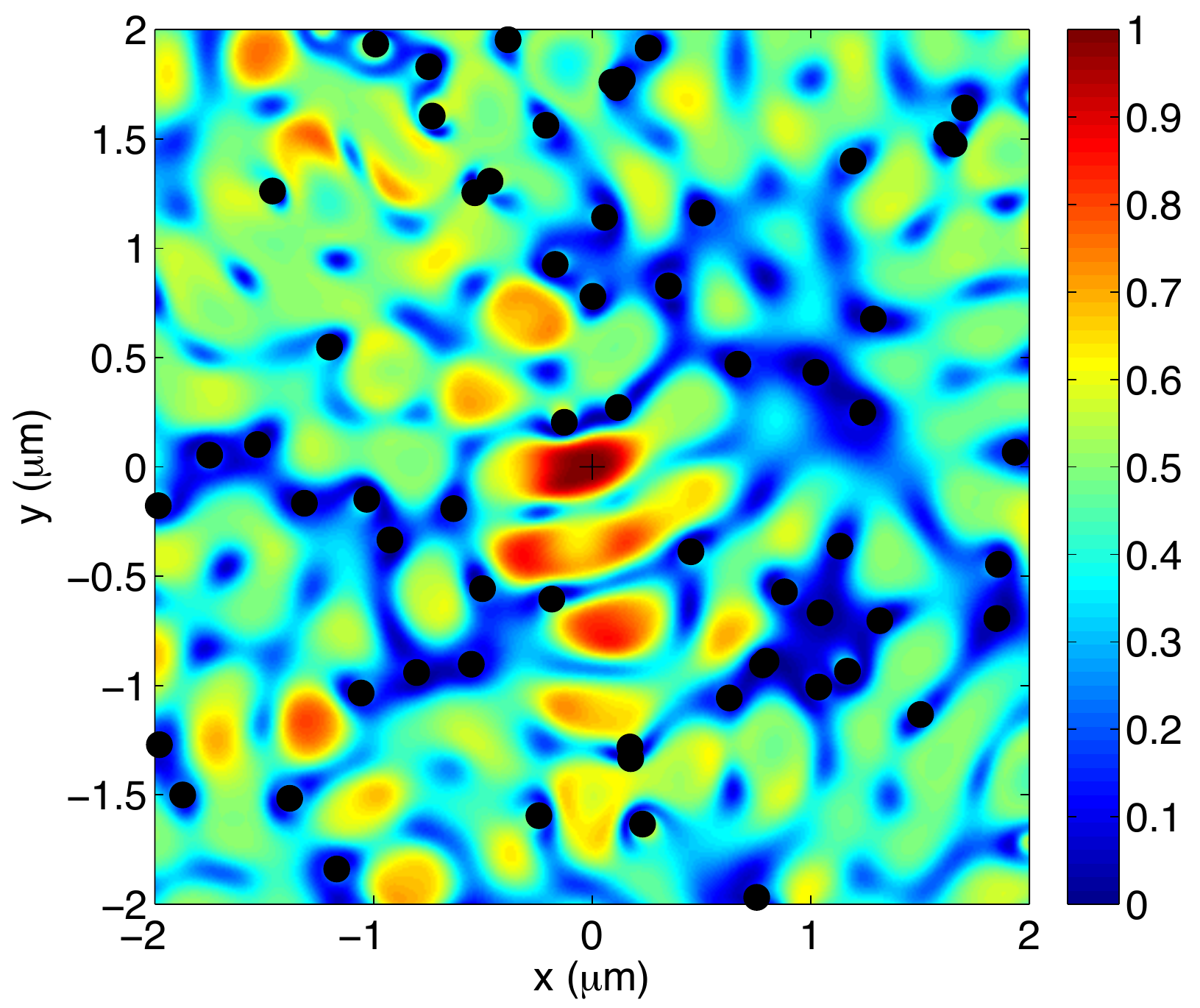}
\vspace{0.cm}
}
\caption{(Color online) Correlation factor $\calG^{(2)}$ plotted as a function of the position of an emitter scanning across the medium, while the other emitter is fixed at the origin (dark cross). Red (resp. blue) colors indicate positions for which superradiant (resp. subradiant) emission is obtained. Black dots show the positions of the scatterers constituting the disordered medium. The scattering cross section and the density of the scatterers are chosen to get multiple scattering in the diffusive regime. Emission wavelength $\lambda=698$ nm.}
 \label{figure2}
\end{figure}

To illustrate these results, we have calculated numerically the correlation factor $\calG^{(2)}$ in a medium structured at the nanoscale, and made of dipole scatterers with random positions. For simplicity, the calculation is performed in two dimensions, for transverse electric polarization and in the diffusive regime (the system is similar to that studied in Ref.~\cite{caze2013strong}, in which details on the numerical approach are given). A map of the correlation factor $\calG^{(2)}$ is shown in Fig.~\ref{figure2}, versus the position of one emitter scanning the image range, while the other emitter is kept at a fixed position at the center.
As expected, $\calG^{(2)}$ varies between 0 and 1 and is maximum at the origin, as the conditions (\ref{G2cond_opt_1}) and (\ref{G2cond_opt_2}) are both fulfilled for emitters very close to each other. We observe large values ($\calG^{(2)}\simeq 0.8$) even for distant emitters, due to the complex underlying photonic mode structure that allows the conditions above to be almost satisfied even at large distances. We also observe low values ($\calG^{(2)}\simeq 0$) in the near field of the scatterers, as the modification of the LDOS they generate yields an unbalanced emissive power, which greatly reduced the possibility of coherent emission. This simple numerical example illustrates the substantial influence of a structured environment on the degree of quantum coherence of the light emitted by two independent single-photon sources.

\emph{Conclusion - }
In summary, we have developed a theoretical framework to describe the influence of a structured environment on the degree of quantum coherence of light emitted by two independent single-photon sources. The analysis provides design rules for the control of the degree of quantum coherence, in terms of classical Green's functions, LDOS and CDOS. In particular, we have established general conditions for the observation of subradiant and superradiant states. The ability of a structured environment to induce long-range coherence, or conversely to inhibit coherent emission even for subwavelength distances between the emitters, has been illustrated numerically on a simple example. These results should have broad applications in the emerging field of quantum nanophotonics, and could suggest new approaches for the sensing of fluorescent sources in complex media.

\subsection*{Acknowledgments}

We acknowledge helpful discussions with A. Goetschy, V. Krachmalnicoff and V. Parigi. This work was supported by LABEX WIFI (Laboratory of
Excellence within the French Program ``Investments for the Future'') under references ANR-10-LABX-24 and ANR-10-IDEX-0001-02 PSL*.

\appendix

\section{CDOS bounded by the LDOS}

In this appendix we demonstrate a Cauchy-Schwarz-like inequality for the CDOS and the LDOS regarding the positions and orientations of the emitters
\begin{align}\label{inequality}
\left| \ImG_{12} \right| \leq \sqrt{ \ImG_{11}} \sqrt{\ImG_{22}}
\end{align}
where $\ImG_{j,k}$ is defined hereinbefore and is proportional to the LDOS for $j=k$ and to the CDOS for $j\neq k$.

We consider two classical dipoles located at $\br_1, \br_2$ with orientations $\bu_1,\bu_2$ and amplitudes $p_1,p_2$ as continuous harmonic sources for the electromagnetic field. The time-averaged power emitted by the two dipoles is a positive quantity that can be written as 
\begin{align}\label{P_p1_p2}
\left< P \right>_T &= \frac{\omega}{2} \Im \left[ \bp_1^* \cdot \bE (\br_1) + \bp_2^* \cdot \bE (\br_2) \right] \nonumber \\
&= \frac{\mu_0 \omega^3}{2} \left[ |p_1|^2 \ImG_{11} + |p_2|^2 \ImG_{22} + 2 \Re \left[ p_1^* p_2 \right] \ImG_{12} \right]
\end{align}
where $\bE$ is the classical electric field generated by the two dipole emitters. In the particular case where the two emitters are in-phase with $p_2=\lambda p_1, \lambda \in \mathds{R}$, the power emitted can be written as
\begin{align}\label{P_lambda}
\left< P \right>_T  = \frac{\mu_0 \omega^3 |p_1|^2}{2} \left[ \ImG_{11} + \lambda^2 \ImG_{22} + 2\lambda \ImG_{12} \right]
\end{align}
which must be a positive quantity for any values of $\lambda$. The determinant of this second order polynomial in $\lambda$ must therefore be negative, which yields
\begin{align}\label{final_inequality}
\left| \ImG_{12} \right| &\leq \sqrt{\ImG_{11} \ImG_{22}} ~ .
\end{align}


\end{document}